\documentclass[12pt]{elsart}
\usepackage{amsmath,amssymb}
\usepackage{graphicx}
\usepackage[bf]{subfigure}
\begin{document}
\begin{frontmatter}

\title{Acoustic phonon transport through a double-bend quantum waveguide}
\author{Yi~Ming\corauthref{cor}},
\author{Zhexian~Wang},
\author{Zejun~Ding\corauthref{cor}}
\corauth[cor]{Corresponding author.} \ead{meanyee@ustc.edu(Yi
Ming)},\ead{zjding@ustc.edu.cn(Zejun Ding)}
\address{Hefei National Laboratory for Physical Sciences at Microscale and\\
Department of Physics, University of Science and Technology of
China\\
Hefei, 230026, Anhui, People's Republic of China}

\begin{abstract}
In this work, using the scattering matrix method, we have
investigated the transmission coefficients and the thermal
conductivity in a double-bend waveguide structure. The transmission
coefficients show strong resonances due to the scattering in the
midsection of a double-bend structure; the positions and the widths
of the resonance peaks are determined by the dimensions of the
midsection of the structure. And the scattering in the double-bend
structure makes the thermal conductivity decreases with the
increasing of the temperature first, then increases after reaches a
minimum. Furthermore, the investigations of the multiple double-bend
structures indicate that the first additional double-bend structure
suppresses the transmission coefficient and the frequency gap
formed; and the additional double-bend structures determine the
numbers of the resonance peaks at the frequency just above the gap
region. These results could be useful for the design of phonon
devices.
\end{abstract}

\end{frontmatter}

\section{\label{sec1}Introduction}

Since the discovery of the quantized electronic conductance
phenomena \cite{ref01,ref02}, the double-bend electron waveguide has
been investigated as a very important case. The problem was firstly
studied by Weisshaar et al. \cite{ref03} using mode-matching
technique. The experiment work on the low temperature conductance of
the double-bend waveguide was carried out by Wu et al. \cite{ref04}
By using the recursive Greens function technique, Kawamura et al.
\cite{ref05,ref06} studied this double-bend waveguide again. All the
works showed strong resonant transmission due to internal
reflections in the special structure and Refs.~\cite{ref05,ref06}
showed the ``existence of an energy gap between the first and second
subband threshold energies where the conductance is suppressed for
multiple double-bend structures''. Based on the property of the
strong resonant transmission through the double-bend waveguide, Shi
et al. \cite{ref07} proposed a simple spin filter with which an
extremely large spin current is expected.

Same as the electronic conductance, the thermal conductivity is also
important for semiconductor nanostructures. For an ideal elastic
beam at an enough low temperature, the thermal conductivity is
dominated by the ballistic phonon and is quantized in a universal
unit, $\pi^2 k_B^2 T/3h$, analogous to the well-known $2e^2/h$
electronic conductance quantum. \cite{ref08,ref09,ref10} These
predictions have been verified experimentally \cite{ref11}. Since
then, many works have been done to study the geometrical effects on
the transmission of the acoustic phonon and the thermal conductivity
through the quantum waveguide with various geometries
\cite{ref12,ref13,ref14,ref15,ref16,ref17,ref18,ref19,ref20}. But as
an important case, the quantum waveguide with a double-bend
structure has not been studied on the phonon transmission
coefficients and thermal conductivity. Therefore, in this work, we
calculate the phonon transmission coefficients and thermal
conductivity in a double-bend quantum waveguide using the
scattering-matrix method \cite{ref21,ref22,ref23,ref24,ref25} and
considering the stress-free boundary condition
\cite{ref15,ref16,ref17,ref18,ref19,ref20}.

The organization of this paper is as follows. In section~\ref{sec2},
we present the model and the numerical method briefly. The numerical
results for a one double-bend structure are discussed in
section~\ref{sec3}. In section~\ref{sec4}, the results of a two
double-bend structures are presented. Finally, a summary is made in
section~\ref{sec5}.

\section{\label{sec2}Model and formalism}

\begin{figure}[htbp]
\begin{center}
\includegraphics[width=0.5\textwidth]{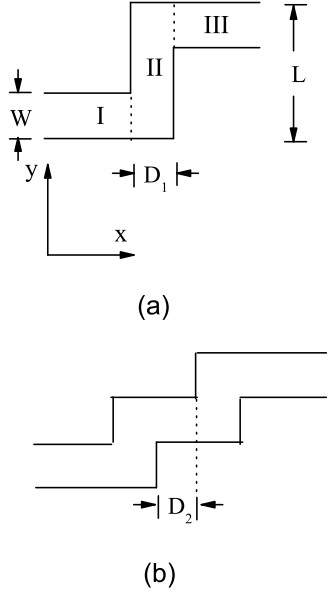}
\end{center}
\caption{\label{fig1} (a) Double-bend waveguide with lead width
$W$, lateral length $L$, and longitudinal length $D_1$. (b) Two
double-bend waveguides in a series with the longitudinal length
$D_2$.}
\end{figure}
The geometry of the double-bend quantum waveguide is sketched in
figure~\ref{fig1}. Regions-I and -III are the leads of the device;
region-II is the midsection. The parameters of this device are $W,
L$ and $D_1$, which are the lateral width of the leads, the lateral
width of the midsection and the longitudinal length of the
midsection, respectively. We assume that the temperature in the two
leads (regions-I and -III) are $T_1$ and $T_2$; and the temperature
difference $\delta T$ ($\delta T=T1-T2>0$) between two leads is very
small. The mean temperature $T$ ($=(T_1+T_2)/2$) may then be adopted
as the temperature of two leads. In this work, we choose the same
thickness for the three regions and let the thickness smaller than
the other two dimensions and also than the wavelength or the
coherence length of the elastic waves; there is no mixing of the $z$
modes and a two-dimensional calculation is then adequate
\cite{ref08,ref09,ref10,ref15}. For the imperfect contact at the
regions-I and -III, the thermal conductance $K$ at temperature $T$
is given by \cite{ref08,ref15}
\begin{equation}\label{eqn1}
K=\frac{\hbar^2}{k_{\mathrm{B}} T^2}\sum_m \frac{1}{2\pi}
\int_{\omega_m}^{\infty}{\tau_m\left(\omega\right) \frac{\omega^2
\e^{\beta
\hbar\omega}}{\left(\e^{\beta\hbar\omega}-1\right)^2}\d\omega},
\end{equation}
where $\tau_m\left(\omega\right)$ is the transmission coefficient
from mode $m$ of region-I at frequency $\omega$ across all the
interfaces into the modes of region-III; $\omega_m$ is the cutoff
frequency of the $m$th mode; $\beta=1/\left(k_{\mathrm{B}}T\right)$,
$k_\mathrm{B}$ is the Boltzmann constant, $T$ is the temperature;
and $\hbar$ is the Plank's constant. It can be seen from
Eq.~(\ref{eqn1}) that the central issue of the problem is to obtain
the transmission coefficient, $\tau_m\left(\omega\right)$.

We consider the scalar model for the elastic wave; the model for
thin geometry at low temperature is used so that the calculation is
two-dimensional. Here, we treat the simplest case for a horizontally
polarized wave (SH) (polarized along the $z$ direction) propagating
in the $x$-direction. So the wave equation of the displacement field
$u$ is
\begin{equation}\label{eqn2}
\frac{\partial^2u}{\partial t^2}-v_\mathrm{SH}^2\nabla^2u=0,
\end{equation}
where $v_\mathrm{SH}$ is related to the mass density $\rho$ and
elastic stiffness constant $C_{44}$ by
\begin{equation}\label{eqn3}
v_{\mathrm{SH}}=\sqrt{C_{44}/\rho}.
\end{equation}
The stress-free boundary conditions at the edges require
$\hat{n}\cdot\nabla{u}=0$, where $\hat{n}$ is the unit vector normal
to the edge. For the double-bend structure depicted in
Fig.~\ref{fig1}, the solution to Eq.~(\ref{eqn2}) in region-$\xi$
(regions I, II and III) can be expressed as
\begin{eqnarray}\label{eqn4}
u^{\xi}\left(x,y\right)=\sum_{m=0}^N\left[A_m^{\xi}\e^{\mathrm{i}k_m^{\xi}
\left(x-x_{\xi}\right)}+B_m^{\xi}\e^{-\mathrm{i}k_m^{\xi} \left(
x-x_\xi\right)} \right]\phi_m^{\xi}\left(y\right),
\end{eqnarray}
where $x_\xi$ is the reference coordinate along the $x$-direction in
region-$\xi$; $k_m^\xi$ is the wavenumber of the transmitted and
reflected waves in region-$\xi$, giving by the energy conservation
condition
\begin{equation}\label{eqn5}
\omega^2={k_m^{\xi}}^2{v_{\mathrm{SH}}^{\xi}}^2+{m^2{\pi}^2
{v_{\mathrm{SH}}^{\xi}}^2}/{W_{\xi}^2},
\end{equation}
with $W_{\xi}$ the transverse dimension of region-$\xi$;
$\phi_m^{\xi}\left(y\right)$ represents the transverse wavefunction
of acoustic mode-$m$ in region-$\xi$,
\begin{eqnarray}\label{eqn6}
\phi_m^{\mathrm{I}}&=&\left\{ \begin{array}{ll}
\sqrt{\frac{2}{W}}\cos{\frac{m\pi}{W}y}  &\left(m \neq 0 \right), \\
\sqrt{\frac{1}{W}}  &\left(m = 0 \right);
\end{array} \right.\label{subeqn1} \\
\phi_m^{\mathrm{II}}&=&\left\{ \begin{array}{ll}
\sqrt{\frac{2}{L}}\cos{\frac{m\pi}{L}y}  &\left(m \neq 0 \right), \\
\sqrt{\frac{1}{L}}  &\left(m = 0 \right);
\end{array} \right.\label{subeqn2} \\
\phi_m^{\mathrm{III}}&=&\left\{ \begin{array}{ll}
\sqrt{\frac{2}{W}}\cos{\frac{m\pi}{W}\left( y-L \right)}  &\left(m \neq 0 \right), \\
\sqrt{\frac{1}{W}}  &\left(m = 0 \right).
\end{array} \right.\label{subeqn3}
\end{eqnarray}
In principle, the sum over $m$ in Eq.~(\ref{eqn4}) includes all
propagating modes and evanescent modes (imaginary $k_m^{\xi}$).
However, in the practical calculation, besides all the propagating
modes we only take a same limited number of evanescent modes into
account to meet the desired precision for each region in a
double-bend structure. The boundary matching conditions require the
continuity of the displacement $u$ and the stress $C_{44}
\partial{u}/\partial{x}$ at the interface of regions-I and -II
and the interface of regions-II and -III. So we can obtain the
equations for the coefficients in Eq.~(\ref{eqn4}). Rewriting the
resulted equations in the form of matrix, we can derive the
transmission coefficient, $\tau_m$, by the scattering matrix method
\cite{ref21,ref22,ref23,ref24,ref25}.

In the calculations, we employ the following values of elastic
stiffness constant and the mass density \cite{ref26}:
$C_{44}(\mathrm{GaAs})=5.99 (10^{10} \mathrm{N}\mathrm{m}^{-2})$ and
$\rho(\mathrm{GaAs})=5317.6 (\mathrm{kg}\mathrm{m}^{-3})$, and
choosing $W=10$~nm.

\section{\label{sec3}Numerical results for a one double-bend structure}

\begin{center}
\begin{figure*}[htbp]
\subfigure[]{
\includegraphics[width=\textwidth]{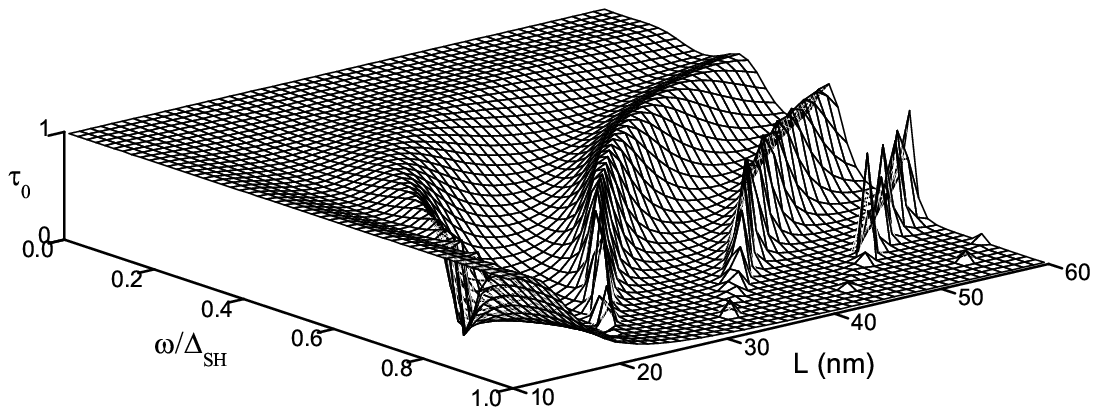}
\label{fig2a}}
\subfigure[]{
\includegraphics[width=\textwidth]{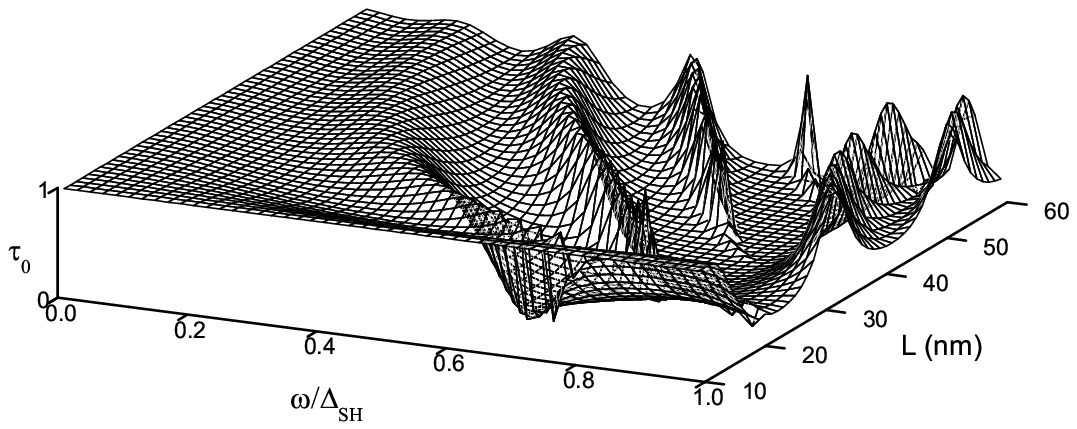}
\label{fig2b}}\caption{\label{fig2} Transmission coefficient
versus the lateral length $L$ and the incident phonon frequency
$\omega/\Delta_{\mathrm{SH}}$ for the double-bend structure with
$W=10$~nm, where $\Delta_{\mathrm{SH}}=\omega_{m+1}-\omega_{m}=\pi
v_{\mathrm{SH}}/W$ ($v_{\mathrm{SH}}$ is the acoustic wave
velocity in GaAs) represents the splitting of the cutoff frequency
between the $(m+1)$th and $m$th mode in the region-I. (a)
$D_1=10$~nm; (b) $D_1=12$~nm.}
\end{figure*}
\end{center}
For acoustic phonon mode, the stress-free boundary condition allows
acoustic waves propagate through the structure in the zero-th mode
with cutoff frequency $\omega=0$. First, we discuss the lowest SH
acoustic mode propagate through the double-bend structure shown in
Fig.~\ref{fig1}(a) from region-I to region-III. The transmission
coefficients of the zero-th mode versus the lateral length $L$ and
the incident phonon frequency $\omega/\Delta_{\mathrm{SH}}$ for the
double-bend structure with $W=10$~nm are shown in Figs.~\ref{fig2a}
($D_1=10$~nm) and \ref{fig2b} ($D_1=12$~nm). One can find from
Fig.~\ref{fig2} that the transmission coefficient approaches unity
as $\omega\rightarrow 0$ and is independent of the lateral length
$L$. This is because, at $\omega \rightarrow 0$, the wavelength of
the phonon is much larger than the dimension of the lateral length
$L$ of region-II and the displacement field $u$ then becomes
essentially the same throughout. We can also find from
Fig.~\ref{fig2} that, for $L=10$~nm, the transmission coefficient is
unity for any incident frequency. In this case the double-bend
waveguide reverts to be a uniform waveguide; and in the uniform
waveguide, the transmission coefficient of the zero-th mode of
phonon is unity. Furthermore, from Fig.~\ref{fig2} one can find more
resonance like features occur in the transmission coefficient with
increasing incident frequency or the lateral length of region-II,
$L$; and the influence of the longitudinal width $D_1$ also appear.

\begin{center}
\begin{figure*}[htbp]
\subfigure[]{
\includegraphics[width=0.4\textwidth]{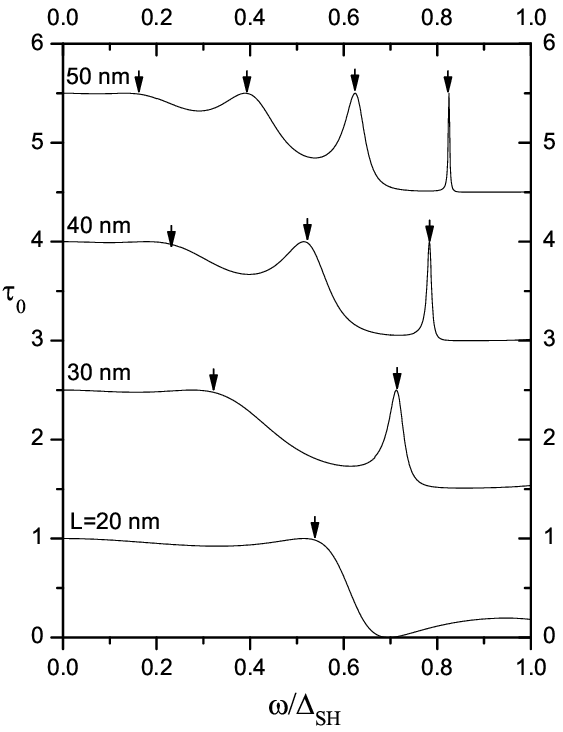}
\label{fig3a}} \subfigure[]{
\includegraphics[width=0.4\textwidth]{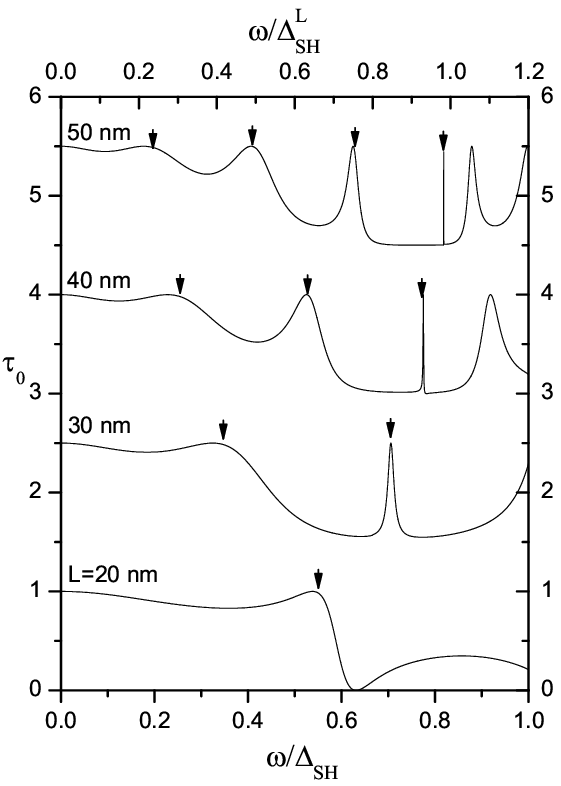}
\label{fig3b}} \subfigure[]{
\includegraphics[width=0.8\textwidth]{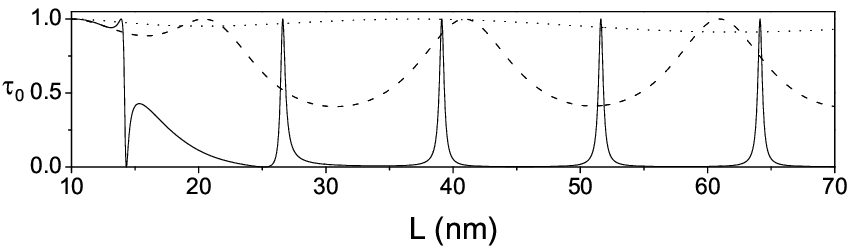}
\label{fig3c}} \subfigure[]{
\includegraphics[width=0.8\textwidth]{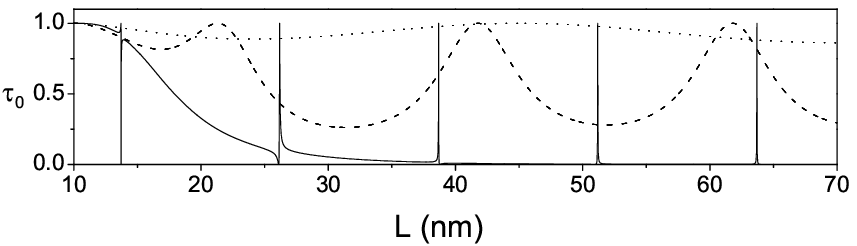}
\label{fig3d}} \caption{\label{fig3} Transmission coefficient as a
function of: (a) the reduced frequency $\omega/\Delta_{\mathrm{SH}}$
with $W=10$~nm and $D_1=10$~nm; (b) the reduced frequency
$\omega/\Delta_{\mathrm{SH}}$ and $\omega/\Delta_{\mathrm{SH}}^L$
with $W=10$~nm and $D_1=12$~nm, where
$\omega/\Delta_{\mathrm{SH}}^L=\pi v_{\mathrm{SH}}/D_1$. Curves are
vertically offset by $1.5$ units for clarity. The arrows indicate
the resonance peaks. Transmission coefficient as a function of the
lateral length $L$ with: (c) $D_1=10$~nm; (d) $D_1=12$~nm. The
dotted, dashed and solid curves in (c) and (d) correspond to
$\omega/\Delta_{\mathrm{SH}}=0.2$, $0.5$ and $0.8$, respectively;
and in (d), these curves correspond to
$\omega/\Delta_{\mathrm{SH}}^L=0.24$, $0.6$ and $0.96$,
respectively.}
\end{figure*}
\end{center}

To clearly see the resonance structure in the transmission
coefficient, the transmission coefficient as a function of
$\omega/\Delta_{\mathrm{SH}}$ for different lateral lengths of
region-II, $L$, and as a function of $L$ for different
$\omega/\Delta_{\mathrm{SH}}$ are shown in Fig.~\ref{fig3}.

In region-II, the propagation wave and reflected wave with the
same wavenumber, $k=\sqrt{\omega^2/v^2-m^2\pi^2/{D_1}^2}$, will
couple together. The reflected wave from region-II will interfere
with the propagation wave in region-I. When the phase shift
between the incident wave in the region-I and the reflected wave
from the region-II reaches $2n\pi$, the transmission will be
enforced, and, $\tau_0=1$. For a fixed lateral length $L$, same as
the resonant transmission of electrons \cite{ref27}, the resonance
condition is approximately $2k_n L=2n\pi$. In Figs.~\ref{fig3a}
and \ref{fig3b}, the arrows mark the resonance positions. In
Fig.~\ref{fig3b}, when $\omega/\Delta_{\mathrm{SH}}^L<1$, the
resonance peak positions are similar as the positions in
Fig.~\ref{fig3a}. This is easy to understand: in this case only
the zero-th mode can exist in region-II, so the wavenumber
$k=\omega/v$ is irrelevant with $D_1$. But for
$\omega/\Delta_{\mathrm{SH}}^L \geq 1$, because of more than one
transverse mode existing in the region-II the transmission
coefficients in Fig.~\ref{fig3b} are totally different with those
in Fig.~\ref{fig3a}. Furthermore, from Figs.~\ref{fig3a} and
\ref{fig3b}, one can find that the closer the frequency to the
cutoff frequency of the $1$th mode in region-II, the narrower the
resonance peak is. This is due to the fact that at the cutoff
frequency of the $1$th mode, the $1$th mode is excited and the
resonant condition is destroyed totally. The transmission
coefficients as a function of the lateral length $L$ show periodic
resonance in Figs.~\ref{fig3c} and \ref{fig3d}. The period is
determined by the condition $2k\Delta L=2\pi$ as, $\Delta
L=\pi/k=\lambda/2$. These results could be useful for the design
of phonon devices.

From the transmission coefficients the thermal conductivity can be
obtained from Eq.~(\ref{eqn1}). Let $\omega=\eta
\Delta_{\mathrm{SH}}$, $T=\zeta
\Delta_{\mathrm{SH}}\hbar/k_\mathrm{B}$ and choose $\hbar=1$ in
Eq.~(\ref{eqn1}), then
\begin{equation}\label{eqn7}
\frac{K}{G_q}=\sum_m \int_{\eta_m}^{\infty} \frac{3\eta^2 \tau_m
\left(\eta \right)}{\pi^2 \zeta^3}
\frac{\mathrm{e}^{\eta/\zeta}}{\left( \mathrm{e}^
{\eta/\zeta}-1\right)^2}\, \mathrm{d} \eta,
\end{equation}
where $G_q=\pi^2 k_{\mathrm{B}}^2 T/3 h$ is the universal value of
the thermal conductivity. Let's consider the integration in above
equation: if $m=10$ and $\zeta \leq 1$, then $\eta_m=10$,
$\mathrm{e}^{\eta/\zeta} \geq \mathrm{e}^{10} \gg 1$, and

\begin{eqnarray}\label{eqn8}
&&\int_{\eta_{10}}^{\infty} \frac{3\eta^2 \tau_{10} \left(\eta
\right)}{\pi^2 \zeta^3} \frac{\mathrm{e}^{\eta/\zeta}}{\left(
\mathrm{e}^ {\eta/\zeta}-1\right)^2}\, \mathrm{d} \eta \nonumber \\
&&\sim \int_{10}^{\infty} \frac{3\eta^2 \tau_{10} \left(\eta
\right)}{\pi^2 \zeta^3} \frac{1}{\mathrm{e}^ {\eta/\zeta}}\,
\mathrm{d} \eta \leq \int_{10}^{\infty} \frac{3\eta^2}{\pi^2
\zeta^3} \frac{1}{\mathrm{e}^ {\eta/\zeta}}\, \mathrm{d} \eta \leq
\int_{10}^{\infty} \frac{3x^2}{\pi^2} \frac{1}{\mathrm{e}^x}\,
\mathrm{d} x \nonumber\\
&&=\frac{3}{\mathrm{e}^{10} \pi^2} \int_{0}^{\infty}
\frac{y^2+20y+100}{\mathrm{e}^y}\, \mathrm{d} y =
\frac{366}{\mathrm{e}^{10} \pi^2} \approx 0.00168,\qquad (\zeta \leq
1), \qquad
\end{eqnarray}
in which we have used the equation,
\begin{eqnarray}\label{eqn9}
\int_{0}^{\infty}\mathrm{e}^{-sx} x^n
\mathrm{d}x=\frac{n!}{s^{n+1}},\qquad\qquad n=0,1,2,\cdots,\quad
s>0.
\end{eqnarray}
Therefore, the sum of $m$ in Eq.~(\ref{eqn1}) can truncated at $m=9$
for $\zeta \leq 1$.

\begin{figure}[htbp]
\begin{center}
\includegraphics[width=\textwidth]{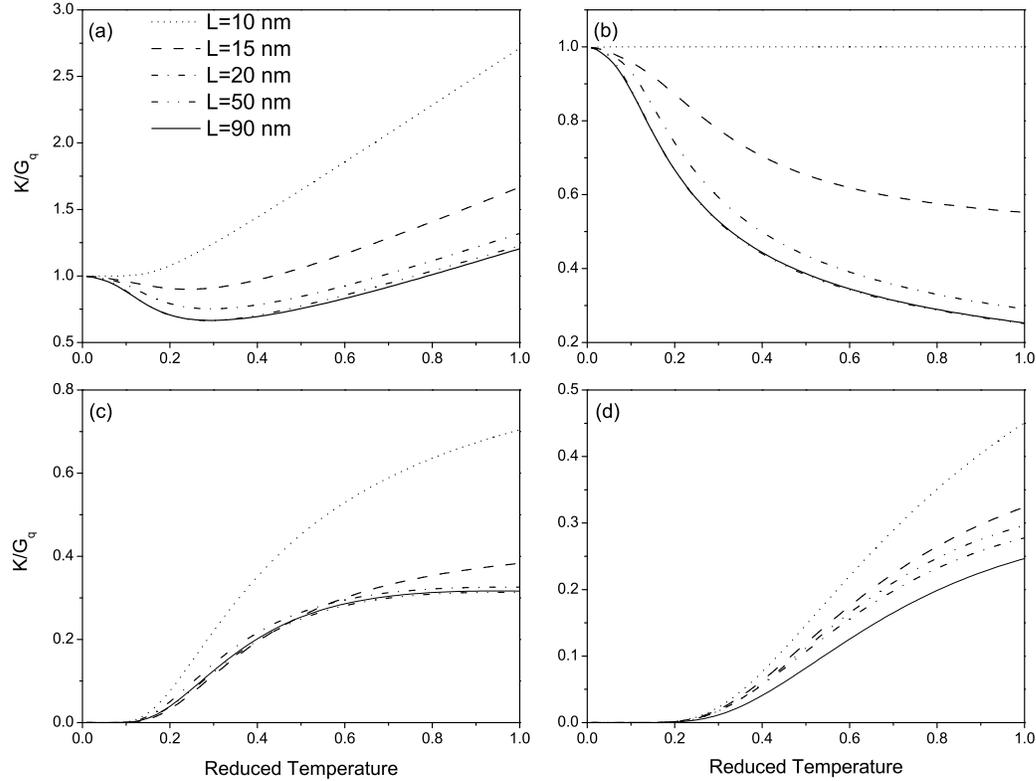}
\end{center}
\caption{\label{fig4} Thermal conductivity reduced by $G_q$, as a
function of the reduced temperature, $\zeta=k_{\mathrm{B}}T/\hbar
\Delta_{\mathrm{SH}}$, for the different lateral lengths of $L$: (a)
the total thermal conductivity; (b), (c) and (d) are the thermal
conductivity of the modes-0, 1 and 2, respectively.}
\end{figure}
\begin{figure}[htbp]
\begin{center}
\includegraphics[width=\textwidth]{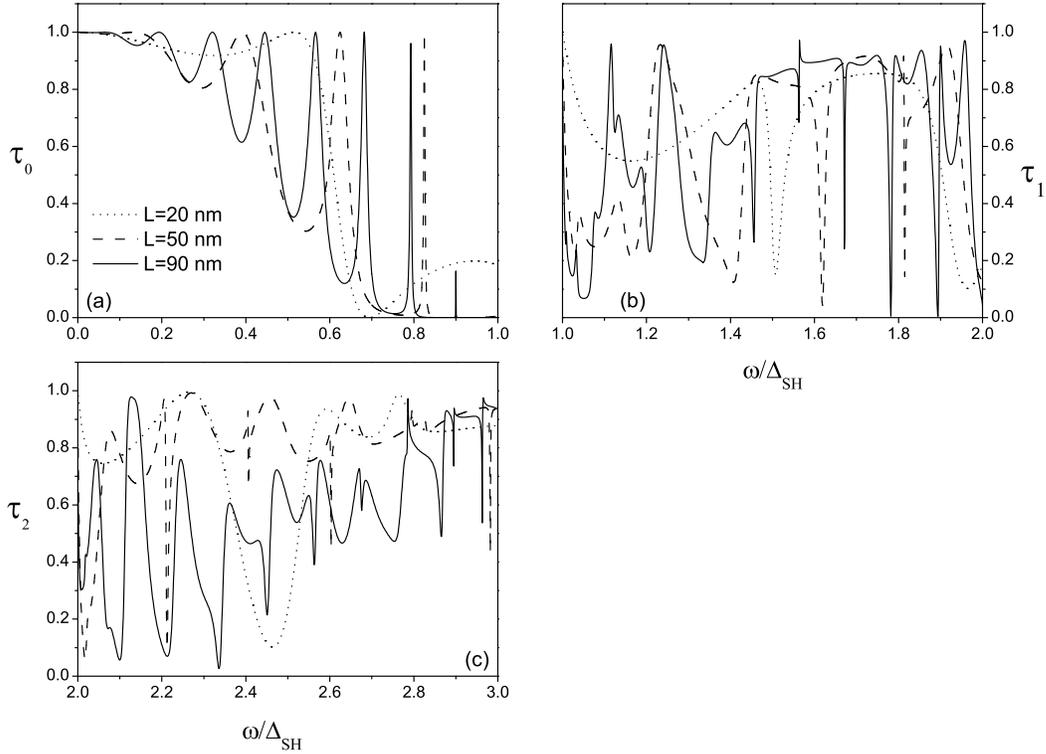}
\end{center}
\caption{\label{fig5} Transmission coefficient as a function of the
reduced frequency $\omega/\Delta_{\mathrm{SH}}$. (a), (b) and (c)
are for $\tau_0$, $\tau_1$ and $\tau_2$, respectively.}
\end{figure}
Fig.~\ref{fig4} shows the reduced thermal conductivity $K/G_q$ as a
function of the reduced temperature $k_{\mathrm{B}}T/\hbar
\Delta_{\mathrm{SH}}$. One can find from Fig.~4(a) that the total
reduced thermal conductivity approaches to the unity (i.e. the
thermal conductivity, $K/T$, approaches to the ideal universal
value, $\pi^2 k_{\mathrm{B}}^2/3h$) when $T\rightarrow 0$; the
result is independent of the lateral length of the double-bend
waveguide. This is because the wavelength of the incident phonons is
much larger than the dimension of the region-II when $T \rightarrow
0$ and, thus, the scattering of phonons in the region-II is
negligible. Fig.~4(a) shows a thermal conductivity plateau at $L=0$
and at low temperatures due to no scattering; and the thermal
conductivity increases lineally with the temperature because the
higher transverse modes are excited at higher temperatures (as shown
in Figs.~4(c) and 4(d)). From Fig.~4(a), for $L\neq 0$, one can find
the thermal conductivity decreases from the universal value to a
minimum at low temperatures, then increases with the the
temperature. This can be explained from Figs.~4(b)-4(d): With
increasing temperature, the thermal conductivity of the mode-0
decreases for $L \neq 0$, so the total thermal conductivity
decreases from the universal value. When the temperature further
increases, the higher mode can be excited and the total thermal will
increase after reach the minimum. Furthermore, Fig.~4(a) shows that
the total thermal conductivity decreases with increasing the lateral
length; and when $L>50$~nm, the total thermal conductivity is
insensitive to further increasing of $L$.

From Fig.~5(a), one can find that the transmission coefficient of
the zero-th mode, $\tau_0$, for $L=20$~nm is higher than that for
$L=50$ and $90$~nm; and $\tau_0$ for $L=50$~nm is similar as for
$L=90$~nm. So for the thermal conductivity of mode-0 shown in
Fig.~4(b), it is higher for $L=20$~nm than for $L=50$ and $90$~nm.
Fig.~5(b) shows the transmission coefficients of the $1$th mode for
different lateral lengths of $L$. Here, $\tau_1$ for $L=90$~nm is
higher than that for $L=50$~nm, so the thermal conductivity for
$L=50$~nm is lower than that for $L=90$~nm in Fig.~4(c). The thermal
conductivity for $L=50$~nm is higher than that for $L=20$~nm shown
in Fig.~4(d), because the transmission coefficient of $2$th mode for
$L=50$~nm is higher than that for $L=20$~nm in Fig.~5(c).

\section{\label{sec4}Numerical results for two double-bend structures}

\begin{center}
\begin{figure*}[htbp]
\subfigure[]{
\includegraphics[width=0.5\textwidth]{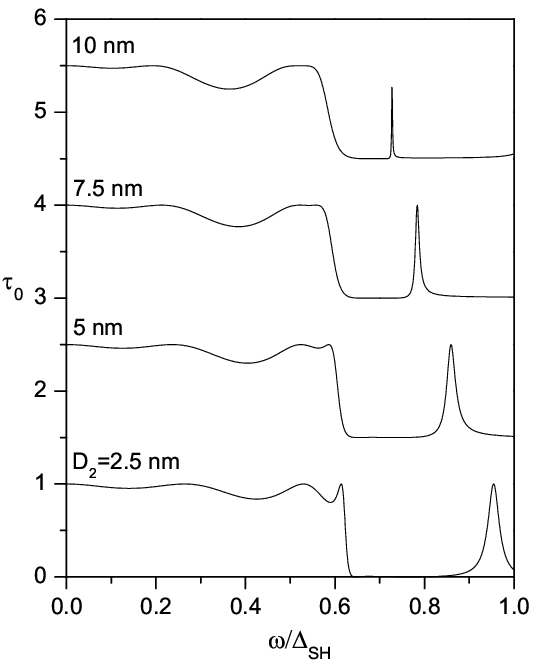}
\label{fig6a}} \subfigure[]{
\includegraphics[width=0.45\textwidth]{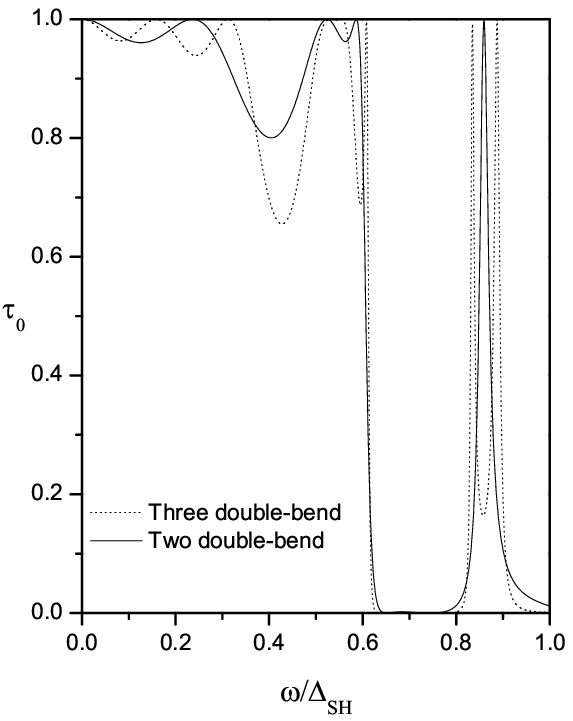}
\label{fig6b}} \caption{\label{fig6} (a), (b) Transmission
coefficient as a function of the reduced frequency
$\omega/\Delta_{\mathrm{SH}}$ for $W=10$~nm, $L=20$~nm and
$D_1=10$~nm. (a) A two double-bend structure for different
longitudinal lengths of $D_2$, where curves are vertically offset by
$1.5$ units for clarity. (b) A two double-bend structure and a three
double-bend structure for the same value of $D_2$.}
\end{figure*}
\end{center}
Fig.~\ref{fig6} shows the transmission coefficient through a
quantum waveguide with more than one double-bend structures.
Fig.~\ref{fig6a} represents the case of a two double-bend
waveguide for different longitudinal lengths of $D_2$. One can
find from the figure that an additional double-bend structure
makes the frequency gap appear; and the gap width decreases with
$D_2$ increases. Furthermore, it is obviously that the width of
resonance peak at the frequency just above the gap region
decreases with $D_2$ increases. Fig.~\ref{fig6b} compares the
transmission coefficients through a two double-bend structures and
a three double-bend structures for the same value of $D_2$. Adding
an additional double-bend structure to the two double-bend
structures doesn't change the frequency gap width but makes an
additional resonance peak appearing at the frequency above the gap
region. It may draw conclusions that the first additional
double-bend structure to the single double-bend structure
suppresses the transmission coefficient and forms a frequency gap;
and an additional resonance peak is formed at the frequency just
above the gap region for each additional double-bend structures.
We can easily expect that an additional double-bend structure
increases the scattering of the phonons so that the thermal
conductivity decreases.

\section{\label{sec5}Summary}

Using the scattering matrix method, we have investigated the phonon
transmission and thermal conductivity in a double-bend waveguide
structure. It is observed strong resonant transmission determined by
the scattering in region-II. The position and the width of the
resonance peaks are determined by the longitudinal length $D_1$ of
region-II. The thermal conductivity approaches to universe value for
$T\rightarrow 0$. This is because the wavelength of the incident
phonons is much larger than the dimension of the region-II when $T
\rightarrow 0$; the scattering of phonons by the region-II is
negligible. Due to the scattering in region-II, the thermal
conductivity of mode-0 decreases with increasing temperature; and at
high temperatures, more than one mode are exited and the thermal
conductivity of these exited modes increases. The total thermal
conductivity decreases thus to a minimum, then increases with the
temperature. The transmission coefficient of a double-bend structure
determines the thermal conductivity for different lateral lengths of
$L$.

In this paper, the transmission coefficients and the thermal
conductivity in multiple double-bend waveguide structures are also
studied. The results show that the first additional double-bend
structure to the single double-bend structure suppresses the
transmission coefficient and forms a frequency gap; and an
additional resonance peak appears at the frequency just above the
gap region for each additional double-bend structure. The additional
double-bend structure suppresses more the thermal conductivity.

\ack{One of authors (Y. Ming) is grateful to Prof.~K.Q. Chen at the
Laboratory of Organic Solids, Center for Molecular Sciences,
Institute of Chemistry, Chinese Academy of Sciences and Dr. W.X. Li
at the Department of Physics, Tsinghua University for helpful
discussions. This work was supported by the National Natural Science
Foundation of China (Grant No.10025420 and 90206009).}

\end{document}